
%
%
\def\pmb#1{\setbox0=\hbox{$#1$}%
\kern-.025em\copy0\kern-\wd0
\kern.05em\copy0\kern-\wd0
\kern-.025em\raise.0433em\box0 }
\def\svec#1{\skew{-2}\vec#1}
\def\Tr{\,{\rm Tr}\,}
\def\ll{\left\langle}
\def\rr{\right\rangle}
\vsize=7.5in
\hsize=5.6in
\tolerance 10000

\baselineskip 12pt plus 1pt minus 1pt
\pageno=0
\centerline{\bf FUSION POTENTIALS FOR $\pmb{G_k}$}
\smallskip
\centerline{{\bf AND HANDLE SQUASHING}\footnote{*}{This
work is supported in part by funds
provided by the U.S. Department of Energy (D.O.E.) under contract
\#DE-AC02-76ER03069, and by the Division of Applied Mathematics of the U.S.
Department of Energy under contract \#DE-FG02-88ER25066.}}
\vskip 24pt
\centerline{Michael Crescimanno}
\vskip 12pt
\centerline{\it Center for Theoretical Physics}
\centerline{\it Laboratory for Nuclear Science}
\centerline{\it and Department of Physics}
\centerline{\it Massachusetts Institute of Technology}
\centerline{\it Cambridge, Massachusetts\ \ 02139\ \ \ U.S.A.}
\vskip 1.5in
\centerline{Submitted to: {\it Nuclear Physics B\/}}
\vfill
\centerline{ Typeset in $\TeX$ by Roger L. Gilson}
\vskip -12pt
\noindent CTP\#2021\hfill October 1991
\eject
\baselineskip 24pt plus 2pt minus 2pt
\centerline{\bf ABSTRACT}
\medskip
Using Chern--Simons gauge theory, we show that the fusion ring of the conformal
field theory $G_k$ ($G$ any Lie algebra) is isomorphic to ${P[u]\over
(\nabla V)}$ where $V$ is a polynomial in $u$ and $(\nabla V)$ is the ideal
generated
by conditions $\nabla V = 0$.  We explicitly construct $V$ for all $G_k$.  We
also derive a residue-like formula for the correlation functions in the
Chern--Simons theory thus providing an RCFT version of the residue formula for
the topological Landau--Ginzburg model.  An operator that acts like a measure
in this residue formula has the interpretation of a handle-squashing operator
and explicit formulae for this operator are given.
\vfill
\eject
\noindent{\bf I.\quad INTRODUCTION}
\medskip
\nobreak
In the study of rational conformal field theories (RCFT), the fusion algebra
plays a central role.  For example, Gepner {\it et al.\/}$^1$ characterized
the fusion rules of $G_k$ Wess--Zumino--Witten (WZW) models and Verlinde$^2$
displayed the connection between modular transformations and the fusion
coefficients.  Over the last few years there has been much progress in better
understanding the fusion rules of $G_k$ theory.$^{3-8,\,32}$
A good introduction to
RCFT and the fusion algebra may be found in Ref.~[9].

Recently, Gepner$^{10}$ has conjectured that the fusion ring $R$ of any RCFT
is isomorphic to ${P[u]\over (\nabla V)}$ where $P[u]$ is a
polynomial ring in $u_i$ over ${\bf Z}$,
the components of some finite-dimensional vector
$u$,  and $(\nabla V)$ is the ideal generated by the
$\nabla V = 0$ where $V$ is a polynomial in the $u_i$'s.  $V$ is called the
fusion potential of the ring $R$.  Gepner was able to show that this
conjecture was true for the RCFT $SU(N)_k$.  In this note we show this
conjecture is true for an arbitrary $G_k$.  Furthermore, the extension of these
techniques to arbitrary coset models appears promising.$^{11}$  Since there is
yet no classification of RCFT's this seems to be about as far as one can come
at the present time to verifying the conjecture.

In a related and somewhat parallel development, Witten$^{12}$ has shown that
Chern--Simons gauge theory in three dimensions is closely connected with
conformal field theory in two dimensions.  This connection has been explored
in many ways.  Understanding the holomorphic quantization and the connection
to the KZ equation$^{13}$ has been studied in Refs.~[14 -- 16] and
understanding the modular transformations of the Hilbert space was detailed
in Refs.~[17,8].  Also, understanding how to explictly
compute fusion algebra via
Chern--Simons theory was studied in Ref.~[8].

In this paper we use Chern--Simons field theory to explicitly construct the
fusion potentials of $G_k$.  We then explore some of the ideas of Ref.~[10] in
the context of Chern--Simons theory to derive an interesting formula for the
correlation functions of Chern--Simons gauge theory.  Perhaps not
surprisingly,
the correlators are given by a residue-like formula, reminiscent of the
correlators of topological models (see Ref.~[18--20,33]).
This formula is
interesting because it relates the measure in Gepner's approach to the
$K$ matrix$^{2,\,21,\,22}$ of a RCFT.

This paper is organized as follows.  Section~II is a short review of the
canonical quantizations of Chern--Simons gauge theory for the case of
three-fold
$T^2\times{\bf R}$ ($T^2$ is the torus).  This technique will be used in
Section~III where we first describe, in general, how to compute the fusion
potential for $G_k$ and then simply write them all down.  Finally, Section~IV
describes correlation functions in Chern--Simons theory from Gepner's point of
view and it is there we encounter a connection between the measure in
Gepner's paper and the $K$ matrix of RCFT.  Section~V is a short conclusion.
\goodbreak
\bigskip
\hangindent=24pt\hangafter=1
\noindent{\bf II.\quad CANONICAL QUANTIZATION OF THE CHERN--SIMONS THEORY ON
$\pmb{T^2\times{\bf R}}$}
\medskip
\nobreak
In Ref.~[12] Witten identified the Hilbert space of Chern--Simons theory as
the space of conformal blocks of the associated RCFT.  We will use this idea
to reduce the computation of the fusion algebra of a $G_k$ conformal field
theory to a quantum
mechanical computation on the Hilbert space of the associated $G_k$
Chern--Simons theory.  In what follows we assume familiarity with
Refs.~[12,14,15,16,17,8] and use the conventions of Ref.~[8] throughout.

To begin, let us recall that we wish to canonically quantize the action
$$I_{\rm CS} = {k\over 4\pi} \int_{\cal N} \Tr \left( A\wedge dA + {2\over 3}
A\wedge A\wedge A\right)\ \ .\eqno(2.1)$$
$A$ is the $g$-valued one-form on ${\cal N}$, $g$ the Lie algebra of $G$,
where ${\cal N}$ will be taken to be the three-fold $\Sigma\times{\bf R}$
($\Sigma$ is a two-dimensional [Riemann] surface), $k$ is, as usual by
gauge invariance, an integer referred to as the level
and $\Tr$ is the symmetric bilinear form of the Lie
algebra $g$ normalized here so that in terms of the generators $\tau^a$ of
the $G$ action $\Tr \left(\tau^a\tau^b\right) = 2\delta^{ab}$.  This is the
Chern--Simons theory related to the $G_k$ RCFT.  To canonically quantize this
action we must first fix a time direction and choose a gauge.  We choose the
time axis to be along the ${\bf R}$ in ${\cal N}$ and choose the axial gauge
$A_t = 0$.  The local coordinates on $\Sigma$ are $x_1$ and
$ x_2$.  Then, as discussed in the above references, we find that this gauge
choice implies the superselection rule (in analogy with electromagnetism it is
called the Gauss' law constraint)
$$F_{12} = \partial_1 A_2 - \partial_2 A_1 + \left[ A_1, A_2\right] = 0
\eqno(2.2)$$
and also the action in this gauge (with $\partial\Sigma=0$) is
$$I_{\rm CS} = {k\over 2\pi} \int_{\cal N} d^3x\Tr \left( A_1 \partial_t
A_2\right)\ \ . \eqno(2.3)$$
Finally the observables of the theory are simply Wilson lines around
non-trivial one-cycles in $\Sigma$,
$${\cal O}_{\mu,c} = \Tr_\mu \left( P\,e^{\int_c A}\right) \eqno(2.4)$$
where $c$ is a one-cycle corresponding to some element of $\pi_1(\Sigma)$,
$\mu$ is a representation of $g$ in which the $\Tr$ is to be taken.

We are not yet ready to quantize the action Eq.~(2.3) because we need to
include factors that come from the measure of the path integral.  As shown for
example in Ref.~[12,15] one can very simply include these factors.  They lead
to an additional term in the Lagrangian and are proportional to the original
action Eq.~(2.1).  Indeed when combined with Eq.~(2.1) they result in simply
shifting $k$ to $k+c$ where $c$ is the quadratic Casimir of $g$,  Thus again
choosing gauge $A_t=0$ and proceeding as before we see that the action we wish
to quantize is simply that of Eq.~(2.3) with $k$ replaced with $k+c$.

For the remainder of the paper we specialize to the case $\Sigma = T^2$.  It
is convenient to first satisfy the constraint Eq.~(2.2) classically and
quantize the remaining degrees of freedom.  Note also that the gauge choice
$A_t = 0$ does not fix the gauge completely:  We may use a time independent
and single-valued gauge transformation to make the gauge field on $T^2$ a
 constant vector field.  Then, classically,
the constraint Eq.~(2.2) implies that the two
components of the vector field must commute in $g$, and so, in general, both
are in the Cartan subalgebra of $g$.  As such define
$$\eqalign{\int_{c1} A &=a_i \nu^i \cr
\int_{c2} A &= b_i \nu^i}\qquad i = 1,2,\ldots,\ \ \quad n = \hbox{Rank}\; G
\eqno(2.5)$$
where $c1$, $c2$ are the cycles associated to the two generators of $\pi_1$
($T^2$), $\nu^i$ are simple roots of $g$,.  We may now plug Eq.~(2.5) into
Eq.~(2.3) with $k$ shifted to $k+c$ and using $(c1,c1)=0=(c2,c2)$, $(c1,c2)=1$
we have
$$I_{\rm CS}= - {k+c\over 2\pi} \int dt\, a_i C^{ij}\partial_t b_j \eqno(2.6)$$
where $C^{ij}= \Tr (\nu^i \nu^j)$.  We have now reduced Chern--Simons theory
in ${\cal N}$ to just a quantum mechanics problem where the $a_i$ and $b_j$
are just the coordinates and the momenta.  The commutator in terms of
$a_i,b_j$ is,
$$\eqalign{\left[ a_i, b_j\right] &= - {2\pi i \over k+c} \left(
C^{-1}\right)_{ij} \cr
\left[ a_i,a_j\right] &= 0 = \left[ b_i, b_j\right] \ \ .\cr}\eqno(2.7)$$
Rather than represent these operators on a Hilbert space we find a more
convenient set of operators is suggested by the observables of the theory
Eq.~(2.4).  Whatever representation  $\mu$ is chosen, ${\cal
O}_{\mu c}$ will involve only exponentials of the $a_i$'s and $b_i$'s.  It is
thus natural to find a Hilbert space realization of
$${\rm A}_j = e^{ia_j}\ \ ,\qquad {\rm B}_j = e^{ib_j}\eqno(2.8)$$
and so Eq.~(2.7) implies
$$\eqalign{
{\rm A}_i {\rm B}_j {\rm A}^{-1}_i {\rm B}^{-1}_j &=
e^{{2\pi i \over k+c}\left( C^{-1}\right)_{ij}} \cr
{\rm A}_i {\rm A}_j &= {\rm A}_j {\rm A}_i \  \ ,
\qquad {\rm B}_i {\rm B}_j = {\rm B}_j {\rm B}_i \ \ .\cr}\eqno(2.9)$$
This is analogous to a Weyl basis for the CCR.  We now simply define a vacuum
$|0\rangle$ by ${\rm A}_i|0\rangle = |0\rangle$ $\forall i$  (note ${\rm A}$,
 ${\rm B}$ are
unitary operators), and from Eq.~(2.9) use the ${\rm B}_i$'s
as raising operators to
generate the Fock space of states.  It is easy to see that due to the fact
that the commutator in Eq.~(2.9) is idempotent, the spectrum of eigenvalues of
the ${\rm A}_i$'s will repeat after some number of applications of the raising
operators ${\rm B}_i$'s.  We may thus consistently truncate this Fock space to
a
finite-dimensional Hilbert space, which we call the ``Hilbert space of the
Gaussian model'' because of the strong resemblance of Eq.~(2.9) with those of
a system of $n$ free bosons.

This resulting finite-dimensional Hilbert space is not yet to be identified
with the space of conformal blocks: the operators ${\rm A}_i$ are not invariant
under the residual gauge invariance associated with Weyl transformations. As
shown in Ref.~[8], under the Weyl action the Gaussian Hilbert space described
above breaks into Weyl covariant subspaces.  Finally, implementing this
remaining gauge invariance we project all the operators of the theory onto the
completely Weyl-odd sector.\footnote{*}{We mean states $\psi$ s.t. $\forall
\omega \in W$ ($W$ is the Weyl group) with $\omega^2=1$, that $\omega\psi
=-\psi$.}  It
is natural from the point of the characters$^{23,\,24,\,17}$ that one should
identify this completely Weyl-odd sector with the conformal blocks.  For more
details see Ref.~[8].

In the basis where ${\rm A}_i|0\rangle = |0\rangle$
$\forall i$ we have finally that
the ${\cal O}_{\mu,c1}$ are diagonal on these completely Weyl-odd states.
Furthermore, because ${\cal O}_{\mu,c2}$ is composed of raising operators and
is, by definition, even under all $\omega \in W$ ($W$ is the Weyl group) it is
easy to see that ${\cal O}_{\mu,c2}$ is a map from completely Weyl-odd states
to themselves.  It is not hard to show that these maps are precisely the
fusion matrices.  In notation, let ${\cal H}$ denote the Hilbert space of
states corresponding to the conformal blocks.  They are labelled by
representations since
$$\psi_0\in{\cal H}\ \ ,\qquad\hbox{and}\qquad \psi_\mu = {\cal
O}_{\mu,c2}\psi_0\ \ .\eqno(2.10)$$
one has
$${\cal O}_{\mu,c2} \psi_\nu = N^\tau_{\mu\nu} \psi_\tau \eqno(2.11)$$
where $N^\tau_{\mu\nu}$ are the fusion coefficients.  The proof of
Eq.~(2.11) is found by comparing the manipulations above to Refs.~[3,7].  Note
the ${\cal O}_{\mu,c1}$ are just the diagonalization of the fusion matrices
${\cal O}_{\mu,c2}$ under $S$;
$${\cal O}_{\mu,c1} = S\,{\cal O}_{\mu,c2} S^{-1} \ \ .\eqno(2.12)$$
This is the Verlinde theorem.$^{2,\,26}$  For more details on the method
described here see Ref.~[8].
\goodbreak
\bigskip
\noindent{\bf III.\quad FUSION POTENTIALS }
\medskip
\nobreak
Having described how one can compute the fusion rules of a $G_k$ RCFT using
Chern--Simons theory, we will now proceed to derive the fusion potential of
any $G_k$.

It is worth mentioning that there are other ways to characterize the fusion
algebra, such as finding the generating functions first introduced in
Refs.~[27,28].  However, in this note we will follow the spirit of Ref.~[10]
where it is conjectured that the fusion ring of any RCFT is isomorphic to
${P[u]\over (\nabla V)}$ where $V(u)$ a polynomial
is called the fusion potential.

We will show that this is true for the conformal field theories $G_k$.  Our
strategy is as follows: We first show that there is a potential  for the
Gaussian model (described in Section~II).  We then demonstrate that the fusion
rules of $G_k$ are realized on a subvariety of the variety ${\cal M}$ defined
by the $\nabla_x V = 0$ conditions of the Gaussian model, and show how the
Weyl action naturally removes from ${\cal M}$ all the points {\it except\/}
those on the subvariety.  Many of the ideas in this section come from
Ref.~[10].  Although the Gaussian model of Eq.~(2.9) is something like a
``free field'' decomposition of the theory (in that it has such a strong
resemblance to a system of $n$ free bosons) this author sees no firm
connection between this approach and that of Ref.~[29].  As the reader will
see below, the idempotency and ``free-field'' Gestalt of Eq.~(2.9) are
the key notions that allow one to integrate the fusion rules to a single
potential.

Before displaying the potentials for $G_k$, we pause to more explicitly
describe the method.  Imagine assigning to each ${\rm A}_i$ operator
of Eq.~(2.8) a
complex variable $ x_i$.  We will find a potential $V(x)$ such that
the ring of fusions in the Gaussian model, given by Eq.~(2.9), will be given by
${P[x]\over \left( \nabla_xV\right)}$.  In accordance with the general
ideas of Gepner,$^{10}$ the solutions $x^{({\svec\ell})}$ of $\nabla_xV=0$
will be an affine variety ${\cal M}$ whose points will be in a $1-1$
correspondence with the states $|{\svec\ell}\rangle$
in the Hilbert space of the Gaussian model, the
$1-1$ map being given by
$$A_i |{\svec\ell}\rangle =  x^{({\svec\ell})}_i |{\svec\ell}\rangle\ \
.\eqno(3.1)$$
We thus see that the variety $\nabla_x V = 0$ is essentially isomorphic to
$\Lambda_W/(k+c)\Lambda_R$.  We wish to ultimately only discuss the ring
of fusions of Weyl-even operators, like the Wilson line of Eq.~(2.4).  Thus
consider the map
$$ x_i \longrightarrow u_i (x)\eqno(3.2)$$
where $u_i$ are invariant under the action of the Weyl group (under the Weyl
group, take the $ x_i$'s to transform as the ${\rm A}_i$'s) and are taken to be
the defining representations of $g$.
Obviously, $P[u]$ contains all the operators
of Eq.~(2.4).  We write the potential of the Gaussian model as a function of
the $u_i$'s.  Then the ring $R = {P[u]\over\left( \nabla_uV\right)}$ is the
fusion ring of $G_k$.  We can show this as follows.  Since
$$\nabla_x V = \left[ {\partial u\over\partial x}\right] \nabla_u V = 0$$
on ${\cal M}$ we ask what subvariety is picked out by $\nabla_u V = 0$.
It is simply ${\cal M}$ minus the points for which $\det\left[ {\partial
u\over\partial x}\right] = 0$.  Now it is simple to show that $\det \left[
{\partial u\over\partial x}\right]$ is completely Weyl-odd and that it is, up
to some trivial factors, the vacuum state (as viewed as an operator)
of the conformal field theory $G_k$
$$\psi_0 \propto \det \left[ {\partial u\over \partial x}\right]_{x_i={\rm
B}_i}
|0\rangle \ \ ,\eqno(3.3)$$
Thus $\det \left[ {\partial u\over\partial x}\right] = 0$ at precisely those
points on ${\cal M}$ that correspond to Weyl orbits of length less than
$|W|$, the
order of the Weyl group.  For those points of ${\cal M}$ on the Weyl orbits
that have length $|W|$ we see that the map Eq.~(3.2) maps all points in
that orbit to the same point.  Thus $\nabla_u V = 0$ corresponds to a
subvariety with a point for each integral representation of $G_k$.  This may
be readily seen by comparing the above construction to the construction of the
space of conformal blocks for $G_k$ described in Ref.~[8].  Indeed,
remembering that the ${\cal O}_{\mu,c1}\big|_{{\rm A}_i = x_i}
\in P[u]$ and writing
Eq.~(2.11) on the variety ${\cal M}$;
$${\cal O}_{\mu,c1} {\cal O}_{\nu,c1} S\,\psi_0 = N^\tau_{\mu\nu} {\cal
O}_{\tau,c1} S\,\psi_0 $$
we see that this equation only gives a condition on the product of polynomials
in $P[u]$ when $\psi_0\not=0$ which, as described above (see Eq.~(3.3)), is
precisely at the points $\nabla_u V=0$.  This shows that $R$, the fusion ring
of $G_k$, is given by by
$$R = {P[u]\over\left(\nabla_uV\right)}\ \ .$$
We next construct the potentials $V$ for $G_k$.  As described above it will be
enough to compute the $V(x)$ of the Gaussian model.  The $ x_i$'s are the
natural variables to write the potential in and are just the $q_i$'s of
Gepner$^{10}$ in the case $G=SU(N)$.  For clarity of exposition we divide
$G$ into two classes; $G$ simply-laced and $G$ non-simply-laced.
\goodbreak
\bigskip
\noindent{\bf $\pmb{G}$ Simply-Laced}
\medskip
\nobreak
If $G$ is simply-laced then the matrix $C^{ij}$ of Eq.~(2.6) is just the
Cartan matrix.  We recall that in a Gaussian fusion ring the ``vanishing''
conditions are just statements of the idempotency of the operators, for
example in $U(1)_k$, $A^{2k} = {\tt1\hskip-.27em l} = B^{2k}$. (Recall that we
have convention that $\Tr (\tau^a\tau^b) = 2\delta^{ab}$.)  It is not
difficult to recognize that idempotency of the operators in Eq.~(2.8) and
Eq.~(2.9) implies the ``vanishing'' conditions (and thereby the fusion rules)
of the $G_k$ theory.  We study idempotency in the Gaussian model of Eq.~(2.9)
in the following way:  We find a list of $n(=\hbox{Rank}\,G)$ linearly
independent vectors ${\svec r}_i$ of ${\bf Z}^n$ with smallest integer
components
such that for each ${\svec r}_i$, $\prod\limits^n_j A^{r_{ij}}_j =
{\tt1\hskip-.27em l}$ (${\tt 1\hskip-.27em l}$ in the Hilbert space of the
Gaussian model).  Since $A_\ell |0\rangle = |0\rangle$ $\forall
\ell=1,\ldots,n$ it is enough (by Schur's lemma) to require $\prod\limits_j
A^{r_{ij}}_j$ commutes with all other operators of the theory, namely we
require
$$\left[ \prod\limits^n_j A^{r_{ij}}_j,B_\ell\right] = 0\qquad \forall \ell,i\
\
.$$
For simply-laced $G$ it is very easy to characterize this set of vectors.  By
virtue of Eq.~(2.9) a possible set of the ${\svec r}_i$ is just given by the
rows of the Cartan matrix $A^i_j$ multiplied by $k+c$,
$$r_{ji} =\left( k+c\right) A^i_j\ \ .\eqno(3.4)$$
Now returning to the notion of the vanishing conditions as specifying a
variety we wish to solve the $n$ simultaneous conditions
$$\prod\limits^n_j  x^{r_{ij}}_j = 1\qquad \forall i=1,\ldots,n\ \ .$$
This may be easily done and here we simply write down the potential whose
gradient $\nabla_xV=0$ are the conditions above.\footnote{*}{Care must be
taken when solving these equations not to remove points or
introduce additional images of the
variety.  See Appendix A.}  Details of the particular case of $SU(N)$ is in
Appendix A, added as an aid to the reader.
\goodbreak
\bigskip
\noindent{$\pmb{\left( A_{N-1}\right)_k = SU(N)_k}$}
\medskip
\nobreak
$(\hbox{Rank} =N-1)$.  The fusion potential is
$$V = { x^{N(k+N)+1}_1\over N(k+N)+1} -  x_1 + \sum\limits^{N-1}_{i=2}
\left( {\alpha^{(k+N+1)}_i\over k+N+1} - \alpha_i\right) \eqno(3.5)$$
in which the $ x_j$'s (that correspond to eigenvalues of the ${\rm A}_j$'s
of Eq.~(2.8)) are
given by $x_j = \alpha_j x^j_1$, $2\le j<N-1$. (Note that the Jacobian
in going from $x_i$'s to $\alpha_i$'s is always non-zero.)
\goodbreak
\bigskip
\noindent{\bf$\pmb{\left( D_\ell\right)_k}$}
\medskip
\nobreak
Let $\kappa=k+c$.  We find that one must distinguish the two cases $\ell=$even
and $\ell=$odd.  Thus we find,
$$V = { x^{R\kappa+1}_1\over R\kappa+1} - x_1 + { x^{R\kappa+1}_2\over
R\kappa + 1} - x_2 + \sum\limits^{\ell-2}_{j=3} \left( {\alpha^{\kappa+1}_j
\over \kappa +1} - \alpha_j\right) \eqno(3.6)$$
where $R=2$ is $\ell$ is even and $R=4$ is $\ell$ is odd.  The other $x_j$,
$3<j\le \ell$ are:
$$ x_j = \alpha_j  x^{-(j-2){\rm mod}\,R}_1  x^{j\,{\rm mod}\,R}_2
\ \ .\eqno(3.7)$$
\goodbreak
\bigskip
\noindent{\bf$\pmb{\left(E_6\right)_k}$}
\medskip
\nobreak
Again let $\kappa = k+c$.  The potential is,
$$V = { x^{\kappa+1}_1\over \kappa+1} -  x_1 + { x^{3\kappa+1}_2\over
3\kappa+1} -  x_2 + \sum\limits^6_{j=3} \left(
{\alpha^{\kappa+1}_j\over\kappa +1} - \alpha_j\right) \eqno(3.8)$$
where the $ x_j\ $ $3\le j \le 6$ are
$$\eqalign{ x_3 &= \alpha_3  x^2_2 \cr
 x_4 &= \alpha_4 \cr}\hskip .4in
\eqalign{ x_5 &= \alpha_5 x_2 \cr
 x_6 &= \alpha_6  x^2_2 \ \ .\cr}\eqno(3.9)$$
\goodbreak
\bigskip
\noindent{$\pmb{\left( E_7\right)_K}$}
\medskip
\nobreak
Let $\kappa = k+c$.  The fusion potential for $\left( E_7\right)_k$ is
$$V = { x^{2\kappa+1}_1\over 2\kappa+1} -  x_1 + \sum\limits^7_{j=2}
\left( {\alpha^{\kappa+1}_j \over\kappa+1} - \alpha_j\right)
\eqno(3.10)$$
where the $ x_j\ $ $2\le j \le 7$ are
$$\eqalign{ x_2 &= \alpha_2\cr
 x_3&=\alpha_3\cr
 x_4 &= \alpha_4\cr}\hskip .4in
\eqalign{
 x_5&= \alpha_5 x_1 \cr
 x_6 &= \alpha_6 \cr
 x_7 &= \alpha_7  x_1 \ \ .\cr} \eqno(3.11)$$
\goodbreak
\bigskip
\noindent{$\pmb{\left(E_8\right)_k}$}
\medskip
\nobreak
We let $\kappa = k+c$.  The fusion potential is,
$$V = { x^{2\kappa+1}_1\over 2\kappa +1} -  x_1 + \sum\limits^8_{j=2}
\left( {\alpha^{\kappa+1}_j\over \kappa+1} - \alpha_j\right)
\eqno(3.12)$$
where the $x_j$,  $2\le j \le 8$ are
$$\eqalign{
x_2 &= \alpha_2\cr
x_3 &= \alpha_3 \cr}\qquad
\eqalign{ x_4 &= \alpha_4 \cr
 x_5 &= \alpha_5 x_1 \cr}\qquad
\eqalign{ x_6 &= \alpha_6\qquad  x_8=\alpha_8\cr
 x_7 &=\alpha_7 x_1\cr} \eqno(3.13)$$
\goodbreak
\bigskip
\noindent{\bf $\pmb{G}$ Non-Simply-Laced}
\medskip
\nobreak
If $G$ is non-simply-laced then the matrix $C^{ij}$ of Eq.~(2.6) will not be
the Cartan matrix.  However, the ``vanishing'' conditions are still a
result of the idempotency of the ${\rm A}_i$'s and ${\rm B}_j$'s
of Eq.~(2.8) and
Eq.~(2.9) and one may modify the argument for the simply-laced case.  Now
the ${\svec r}_i$ vectors will correspond to rows in the matrix $C^{ij}$.
$$r_{ji} = (k+c) C^{ji}\ \ .\eqno(3.14)$$
We will call $k+c = \kappa$ throughout.  We will now write down the $C^{ji}$
matrix for each non-simply-laced group and also write down the corresponding
potential for $G_k$.
\goodbreak
\bigskip
\noindent{\bf$\pmb{\left( B_\ell\right)_k}\quad (\ell\ge2)$}
\medskip
\nobreak
The $C^{ij}$ matrix is $(\ell\times\ell)$,
$$\left[ \matrix{
\phantom{-}4 & -2 & & & \cr\noalign{\vskip 0.2cm}
-2 & \phantom{-}4 & -2 & & \cr\noalign{\vskip 0.2cm}
& -2 & \ddots & & \cr\noalign{\vskip 0.2cm}
& & & \phantom{-}4 & -2\cr\noalign{\vskip 0.2cm}
& & & -2 & \phantom{-}2 \cr}\right] \eqno(3.15)$$
and the corresponding potential is,
$$V = { x^{2(\ell-1)\kappa+1}_1\over 2(\ell-1) \kappa+1} -  x_1 +
\sum\limits^\ell_{i=2} \left( {\alpha^{2\kappa+1}_j\over 2\kappa+1} -
\alpha_j\right) \eqno(3.16)$$
where the $ x_j$, $2\le j \le \ell$ are given as $ x_j =  x^j_1$.
\goodbreak
\bigskip
\noindent{\bf$\pmb{\left( C_\ell\right)_k}\quad (\ell\ge3)$}
\medskip
\nobreak
The $C^{ij}$ matrix is $(\ell\times\ell)$,
$$\left[ \matrix{\phantom{-}2 & -1 & & & \cr\noalign{\vskip 0.2cm}
-1 & \phantom{-}2 & -1 & & \cr\noalign{\vskip 0.2cm}
& -1 & \ddots & & \cr\noalign{\vskip 0.2cm}
& & & \phantom{-}2 & -2 \cr\noalign{\vskip 0.2cm}
& & & -2 & \phantom{-}4 \cr}\right]\ \ .\eqno(3.17)$$
In writing the potential down it is convenient to distinguish the cases $\ell$
even and $\ell$ odd.  We find:
$$\eqalignno{
\ell\ \hbox{even:}\qquad V &= { x^{2\kappa+1}_1\over 2\kappa+1} -  x_1 +
{ x^{2\kappa+1}_\ell\over 2\kappa+1} -  x_\ell +
\sum\limits^{\ell-1}_{j=2} \left( {\alpha^{\kappa+1}_j\over \kappa+1} -
\alpha_j\right) &(3.18) \cr\noalign{\hbox{where for $2\le j \le \ell-1$, $ x_j
= \alpha  x^{j\,{\rm mod}\,2}_1$ and}}
\ell\ \hbox{odd:}\qquad V &= { x^{4\kappa+1}_\ell\over 4\kappa+1} -
 x_\ell + \sum\limits^{\ell-1}_{i=1} \left( {\alpha^{\kappa+1}_j\over
\kappa+1} - \alpha_j\right) &(3.19) \cr}$$
where $ x_1 = \alpha_1  x^2_\ell$ and $ x_j = \alpha_j  x^{j\,{\rm
mod}\,2}_2$ for $2\le j \le \ell-1$.
\goodbreak
\bigskip
\noindent{\bf$\pmb{\left( F_4\right)_k}$}
\medskip
\nobreak
The $C^{ij}$ matrix is,
$$\left[ \matrix{
\phantom{-} 2 & -1 & & \cr\noalign{\vskip 0.2cm}
-1 & \phantom{-}2 & -2 & \cr\noalign{\vskip 0.2cm}
& -2 & \phantom{-}4 & -2 \cr\noalign{\vskip 0.2cm}
& & -2 & \phantom{-}4 \cr}\right] \eqno(3.20)$$
and the corresponding potential is
$$V = { x^{\kappa+1}_1\over\kappa+1} -  x_i + { x^{\kappa+1}_2\over
\kappa+1} -  x_2 + { x^{\kappa+1}_3\over 2\kappa+1} -  x_3 +
{ x^{2\kappa+1}_4\over 2\kappa+1} -  x_4 \ \ .\eqno(3.21)$$
\goodbreak
\bigskip
\noindent{\bf$\pmb{\left( G_2\right)_k}$}
\medskip
\nobreak
The $C^{ij}$ matrix is,
$$\left[ \matrix{\phantom{-}2 & -3 \cr\noalign{\vskip 0.2cm}
-3 & \phantom{-}6 \cr}\right]\eqno(3.22)$$
and the potential is,
$$V = { x^{\kappa+1}_1\over \kappa+1} -  x_1 + { x^{3\kappa+1}_2\over
3\kappa+1} -  x_2\ \ .\eqno(3.23)$$
\goodbreak
\bigskip
\noindent{\bf IV.\quad CORRELATION FUNCTIONS AND HANDLE-SQUASHING}
\medskip
\nobreak
In this section we use the ideas of Ref.~[10] and some elementary facts about
the Gaussian model to suggest an interesting connection between the measure
(for inner products and correlators) used by Gepner$^{10}$ and the $K$ matrix
of Verlinde$^{2,\,34}$ (see Bott$^{22}$).  The Gaussian model
will suggest a simple formula for $K^{-1}$.

We begin by noting that all the potentials of the last section look simply
like the potential of a theory made by tensoring some number of ``free
fields.''  The many Maxwell conditions one would have if one tried to combine
the vanishing conditions of $G_k$ into one potential might
look very restrictive but
in these ``free field'' variables the Maxwell conditions are trivial and have
no content.  Indeed, integrating the ``vanishing'' conditions
seems artificial and only
serves to make contact with what was done for Landau--Ginzburg
models:$^{19,\,20,\,30}$ integration and differentiation of a ${\bf C}$-valued
variable $ x_i$ with respect to $ x_i$ makes sense but (thinking of the map
Eq.~(3.2) above) such an integration or differentiation in the space of
integrable representations does not seem to have a natural interpretation in
the conformal field theory.\footnote{*}{Note that the Jacobian of partials of
the map Eq.~(3.2) does have the loose interpretation as a map from the Wilson
line representations to the space of states in $G_k$.}  So, instead of
proceeding as Gepner$^{10}$ has by defining an inner product (and thereby
correlators) as integrations of polynomials with respect to some measure, we
seek a way of defining the inner product that is more natural {\it
vis-a-vis\/} the conformal field theory.

To motivate our method recall that in the variety $\nabla_u V=0$ each point
corresponds to an integrable representation (and therefore to a state) of the
rational conformal field theory.  Furthermore, by Eq.~(3.1) each point's
$ x^{(\ell)}_i$-value is the entry of the matrix $A_i$ along the diagonal
in the $(\ell,\ell)^{\rm th}$ positions.  Also recall that the inner product
on the Hilbert space of $G_k$ really came from the inner product on the
Hilbert space of the Gaussian model, the norm of which was set by $\ll
0|0\rr=1$.  Finally, we know that in terms of the raising operators (the
${\rm B}_i$'s) of the Gaussian model there exists a distinguished polynomial
$\Gamma({\rm B})$
such that the vacuum state $\psi_0$ of the $G_k$ may be written as,
$$\psi_0 = \Gamma({\rm B}) |0\rangle\ \ .\eqno(4.1)$$
Now combining all these ideas we have a description of the inner product in
terms of the operators of the conformal field theory,
$$\eqalign{\delta_{ij} &= \left( \psi_i, \psi_j\right) = \left( \psi_0, {\cal
O}_{\bar{i},c2} {\cal O}_{j,c2} \psi_0\right) \cr
&= \ll 0 \left| \Gamma^+({\rm B})
{\cal O}_{\bar{i},c2} {\cal O}_{j,c2} \Gamma({\rm B}) \right|
0 \rr \cr
&= \ll 0 \left| S^+ {\cal O}_{\bar{i},c1} {\cal O}_{j,c1} \Gamma^+({\rm A})
\Gamma({\rm A})
S\right|0\rr\ \ .\cr}\eqno(4.2)$$
Now, for a Gaussian model $S|0\rangle= {1\over
\sqrt{R}}\sum\limits^R_\ell
|\ell\rangle$ (sum runs over all states of the Gaussian model) and thus
$$\left( \psi_i, \psi_j\right) =\delta_{ij} = {|W|\over R} \Tr_{\cal H} \left[
{\cal O}_{\bar{i},c1} {\cal O}_{j,c1} \Gamma^+ ({\rm A}) \Gamma({\rm A})
\right] \eqno(4.3)$$
where the trace is taken over, ${\cal H}$, just the states in $G_k$ (the
$\Gamma$'s project out everything in the Gaussian Hilbert space {\it except\/}
the states in $G_k$).  $|W|$ is the order of the Weyl group and $R$ is the
total number of states in the Gaussian Hilbert space, {\it i.e.\/}
$$R = \left| {\Lambda_W\over (k+c)\Lambda_R}\right|\ \ .$$
Further, we may write this as
$$\left( \psi_i, \psi_j\right) = \delta_{ij} = {|W|\over R}
\sum\limits_{\nabla_uV=0} \left( {\cal O}_{\bar{i}}
{\cal O}_j \Gamma^+\Gamma\right)
(x) \eqno(4.4)$$
where the sum is over the variety $\nabla_uV=0$.  Thus ``integration'' may be
understood as a sum of the values of a function evaluated on the points of the
variety.  (The ${\cal O}_i$ and ${\cal O}_j$ in Eq.~(4.4) are ``polynomials''
of Gepner as described in Section~III).  Equation~(4.4) bears a striking
resemblance to the residue formulae of Refs.~[18 -- 20] and is, in a sense, the
RCFT version of those formulae.  Note that this is a genus one formula and its
generalization to the space of operators in
higher genus is not immediately obvious.

Note that although $\Gamma$ is completely Weyl-odd (see Eq.~(4.1)) the
operator $\Gamma^+\Gamma$ is Weyl-even.  $\Gamma^+\Gamma$ is also positive and
real.  Furthermore, by construction
$$\left[ \Gamma^+\Gamma({\rm B}), {\cal O}_{\mu,c2}\right] = 0 \qquad \forall
\mu
\ \ .\eqno(4.5)$$
$\Gamma^+\Gamma$ is the conformal field theory analogue of Gepner's
measure.$^{10}$  It may be expressed as a vector in the space of the operators
${\cal O}_\mu$.  This follows by virtue of Eq.~(4.3).

It is not difficult to show that ${|W|\over R} \Gamma^+\Gamma = K^{-1}$,  the
handle squashing operator.$^{23,\,33,\,34}$  For example, Fig.~1 is a
diagrammatic picture of
Eq.~(4.3).  One may prove ${|W|\over R} \Gamma^+\Gamma = K^{-1}$ directly by
using the Verlinde formula Eq.~(2.12).  Indeed suppose there exists an ${M}$
such that
$$\left[ {\cal O}_{\mu,c1},M\right] = 0 \qquad \forall \mu\eqno(4.6)$$
and,
$$\delta_{ij} = \Tr_{\cal H} \left( {\cal O}_{\bar{i},c1}\,M\,
{\cal O}_{j,c1}\right)
\ \ .$$
By Eq.~(4.6) such an $M$ defines a Hermitian  bilinear form on the operators
${\cal O}_{\mu,c1}$  Thus we have
$$\eqalign{\delta_{ij} &= \sum\limits_\ell \left( \psi_\ell\left| {\cal
O}_{\bar{i},c1} {\cal O}_{k,c1} M \right| \psi_\ell\right) \cr
&= \sum\limits_\ell N^m_{\bar{i}j}
\left( \psi_\ell\left| {\cal O}_{m,c1} M \right|
\psi_\ell\right) \cr}\eqno(4.8)$$
and using
$${\cal O}_{i{\rm cl}} \psi_j = {S^{+j}_i\over S^j_0} \psi_j$$
and Eq.~(2.12) we find that
$$\left( \psi_\ell |M| \psi_\ell\right) = M_{\ell\ell} =
\left|S^\ell_0\right|^2\ \ .$$
Thus $MK = {\tt1\hskip-.27em l}$.

One final remark is in order.  We can use the above descriptions of $G_k$ to
give an expression for $K^{-1}$.
{}From the method described in Section~II it is clear that
$\Gamma({\rm B})$
is simply characterizable as the operator in the Gaussian model
that is associated to the completely Weyl-odd state ($\psi_0$ the vacuum of
$G_k$, see Eq.~(4.1)) containing the
vector $|\rho\rangle$ where $\rho={1\over 2}
\sum\limits_{\alpha\in\Delta^+}\alpha$.  This gives one an explicit way of
computing $K^{-1}$ in the Hilbert space of $G_k$.
As an example we give the following formulae for the case
$G = SU(2)$ and $G=SU(3)$;
$$\eqalign{G &= SU(2) \left\{\eqalign{\Gamma({\rm B}) &= {{\rm B}-{\rm B}^{-1}
\over \sqrt{2}} \cr
K^{-1} &= {1\over 2(k+2)} \left( 3{\cal O}_1-{\cal O}_3\right) \cr}\right.
\cr\noalign{\vskip 0.3cm}
G&= SU(3) \left\{ \eqalign{ \Gamma({\bf B}) &={1\over \sqrt{6}}\left(
{\rm B}_1 {\rm B}_2 -
{\rm B}^{-1}_1 {\rm B}^2_2 + {\rm B}^{-2}_1 {\rm B}_2 - {\rm B}^2_1
{\rm B}^{-1}_2 + {\rm B}_1 {\rm B}^{-2}_2 - {\rm B}^{-1}_1
{\rm B}^{-1}_2\right) \cr
K^{-1} &= \left| {\Lambda_W\over (k+3)\Lambda_r}\right|^{-1} \left( 9{\cal
O}_1 - 6 {\cal O}_8 + 3\left( {\cal O}_{10} + {\cal O}_{\overline{10}}\right)
- {\cal O}_{27}\right)\cr}\right.\cr}\eqno(4.9)$$
These formulae are true for all $k$.  It is intriguing that $K^{-1}$ has this
universal description.
\goodbreak
\bigskip
\noindent{\bf V.\quad CONCLUSIONS }
\medskip
\nobreak
In this note we have shown that there is a fusion potential for all $G_k$ and
have used the ideas of Ref.~[10] and Ref.~[8] to motivate a rather explicit
description of the
handle squashing operator $K^{-1}$.

In closing we note that many of these notions seem to allow simple
generalizations to coset models$^{11}$ and that a more group-theoretic
approach to the fusion potentials is being pursued by Schnitzer,$^{31}$ and
that recently there has been progress in studying the fusion rules from the
$N=2$ Landau--Ginzburg approach.$^{35}$
\vfill
\eject
\centerline{\bf REFERENCES }
\medskip
\item{1.}D. Gepner and E. Witten, {\it Nucl. Phys.\/} {\bf B278} (1986) 493.
\medskip
\item{2.}E. Verlinde, {\it Nucl. Phys.\/} {\bf B300} (1988) 389.
\medskip
\item{3.}M. A. Walton, {\it Nucl. Phys.\/} {\bf B340} (1990) 777.
\medskip
\item{4.}M. Spiegelglas, {\it Phys. Lett.\/} {\bf B247} (1990) 36.
\medskip
\item{5.}M. Spiegelglas, {\it Phys. Lett.\/} {\bf B245} (1990) 169.
\medskip
\item{6.}V. G. Ka\v{c}, talk presented at the {\it
Canadian Mathematical Society
Meeting on Lie Algebras and Lie Groups\/} (University of Montr\'eal, August
1989).
\medskip
\item{7.}J. Fuchs and P. van~Driel, {\it Nucl. Phys.\/} {\bf B346} (1990) 632.
\medskip
\item{8.}M. Crescimanno and S. A. Hotes, ``Monopoles, Modular Invariance and
Chern--Simons Field Theory,'' preprint UCB-PTH-90/38.
\medskip
\item{9.}P. Ginsparg, in {\it Les Houches Lectures\/} (1988).
\medskip
\item{10.}D. Gepner, ``Fusion Rings and Geometry,'' preprint NSF-ITP-90-184.
\medskip
\item{11.}M. Crescimanno and S. A. Hotes, in preparation.
\medskip
\item{12.}E. Witten, {\it Commun. Math. Phys.\/} {\bf 121} (1989) 351.
\medskip
\item{13.}V. Knizhnik and A. B. Zamolodchikov, {\it Nucl. Phys.\/} {\bf B247}
(1984) 83.
\medskip
\item{14.}M. Bos and V. P. Nair, {\it Int. Journ. Mod. Phys.\/} {\bf A5}
(1990) 959; {\it Phys. Lett.\/} {\bf B223} (1989) 61.
\medskip
\item{15.}S. Elitzer, G. Moore, A. Schwimmer and N. Seiberg, {\it Nucl.
Phys.\/} {\bf B326} (1989) 108.
\medskip
\item{16.}A. P. Polychronakos, {\it Ann. Phys.\/} {\bf 203} (1990) 231.
\medskip
\item{17.}T. R. Ramadas, I. M. Singer and J. Weitsman, {\it Commun. Math.
Phys.\/} {\bf 126} (1989) 409.
\medskip
\item{18.}R. Dijkgraff, H. Verlinde and E. Verlinde, {\it Nucl. Phys.\/} {\bf
352} (1991) 59.
\medskip
\item{19.}C. Vafa, {\it Mod. Phys. Lett. A\/} {\bf 6} (1991) 337.
\medskip
\item{20.}K. Intrilligator, ``Fusion Residues,'' Harvard University
 preprint HUTP-91/A041.
\medskip
\item{21.}H. Verlinde and E. Verlinde, ``Conformal Field Theory and Geometric
Quantization,'' Princeton University preprint PUPT-89/1149.
\medskip
\item{22.}R. Bott, {\it Surveys in Diff. Geom.\/} {\bf 1} (1991) 1.
\medskip
\item{23.}V. G. Ka\v{c} and D. H. Peterson, {\it Adv. in Math.\/} {\bf 53}
(1984) 125.
\medskip
\item{24.}V. G. Ka\v{c} and M. Wakimoto, {\it Ad. in Math.\/} {\bf 70} (1988)
156.
\medskip
\item{25.}C. Imbimbo, {\it Phys. Lett.\/} {\bf B258} (1991) 353.
\medskip
\item{26.}G. Moore and N. Seiberg, {\it Phys. Lett.\/} {\bf B212} (1988) 451.
\medskip
\item{27.}C. J. Cummins, P. Mathieu and M. A. Walton, {\it Phys. Lett.\/} {\bf
B254} (1991) 386.
\medskip
\item{28.}L. B\'egin, P. Mathieu and M. A. Walton, ``New Example for a
Generating Function for WZWN Fusion Rules,'' preprint LAVAL-PHY-23/91.
\medskip
\item{29.}D. Gepner and R. Cohen, {\it Mod. Phys. Lett.\/} {\bf A6} (1991)
2249.
\medskip
\item{30.}M. Spiegelglas, ``Setting Fusion Rings in Topological
Landau--Ginzburg,'' Technion preprint PH-8-91.
\medskip
\item{31.}H. Schnitzer, private communication.
\medskip
\item{32.}M. Bourdeau, E. Mlawer, H. Riggs and H. Schnitzer, ``The
Quasirotational Fusion Structure of $SO(M|N)$ Chern--Simons and QZW
Theories,'' Brandeis preprint BRX--TH--319.
\medskip
\item{33.}E. Witten, {\it Nucl. Phys.\/} {\bf B340} (1990) 281.
\medskip
\item{34.}E. Verlinde and H. Verlinde, ``Conformal Field Theory and Geometric
Quantization,'' Princeton preprint PUPTH--89--1149.
\medskip
\item{35.}D. Nemeschansky and N. P. Warner, ``Topological Matter, Integrable
Models and Fusion Rings,'' USC preprint USC-91/031.
\vfill
\eject
\centerline{\bf APPENDIX A}
\bigskip
In this Appendix we describe how one solves the conditions $\prod\limits^n_j
 x^{r_{ji}}_j = 1$.  The ${\svec r}_i$'s correspond to rows in some matrix,
$C^{ij}$.  A little thought indicates that the algebraic manipulations used in
solving $\prod\limits^n_{j}  x^{r_{ji}}_j=1$ correspond to finding a matrix
$a_\ell{}^m$ with $a_\ell{}^m\in{\bf Z}$ and $\det \left|a_\ell{}^m\right|=1$
such that a new $C^{ij}$ defined through
$$C_{\rm new}{}^{ij} = C^{i\ell} a_\ell{}^j\eqno(\hbox{A}.1)$$
has a simpler form than $C^{i\ell}$.  That is, if ${\svec r}_i$'s are taken
to be the rows of $C_{\rm new}{}^{ij}$ then we wish to find $a_\ell{}^j$'s such
that in these new ${\svec r}_i$'s the conditions $\prod\limits^n_j
 x^{r_{ji}}_j=1$ all involve at most two different $ x_i$'s each.  Note
the conditions that $a_\ell{}^m\in{\bf Z}$ and $\det \left|
a_\ell{}^m\right|=1$ are just the requirement that the new system (in the
${\svec r}_s$'s taken as rows in $C_{\rm new}{}^{ij}$) has neither added to nor
removed points from the variety specified by the original set of conditions.
An
$a_\ell{}^m$ may very simply be found for all the $C^{ij}$'s in the text by
careful row reductions $C^{ij}$.

For example, for $SU(N)$ it is not difficult to show that with an
$$a_i{}^\ell =\left[ \matrix{ 1 &x&x'&\ldots& \cr\noalign{\vskip 0.2cm}
0& 1 &x'&\ddots\cr\noalign{\vskip 0.2cm}
0&0& 1 &  & \cr\noalign{\vskip 0.2cm}
\vdots&\ddots & & \ddots & \cr\noalign{\vskip 0.2cm}
 &&&&1\cr}\right] \qquad (N-1)\times (N-1) \eqno(\hbox{A}.2)$$
(where the $x$'s means some integers) that the $C^{ij}_{\rm new}$ ($=$ the
Cartan
matrix for the case of $SU(N)$) may be put into the form
$$C^{ij}_{\rm new} = \left[ \matrix{
2 & -1 \cr\noalign{\vskip 0.2cm}
&&&\ddots && \vdots & \vdots \cr\noalign{\vskip 0.2cm}
&&&&0 & -1 & \phantom{-}0 & \phantom{-}0 \cr\noalign{\vskip 0.2cm}
&&&&&\phantom{-}0&-1&\phantom{-}0\cr\noalign{\vskip 0.2cm}
N-1&&&&&&\phantom{-}0&-1\cr\noalign{\vskip 0.2cm}
N & \phantom{-}0 & \phantom{-}0 & \ldots &&&&\phantom{-}0
\cr}\right]\eqno(\hbox{A}.3)$$
from which follows the potential Eq.~(3.5).
\vfill
\eject
\centerline{\bf FIGURE CAPTION}
\vskip 2in
\centerline{Fig.~1:\quad
The operator ${|W|\over R}\Gamma^+\Gamma$ is $K^{-1}$, the
handle-squashing operator.}
\par
\vfill
\end